\documentclass[runningheads]{llncs}
\usepackage{graphicx}
\usepackage{xcolor}
\pdfoutput=1

\begin{document}

\title{Human-Computer Interaction Considerations When Developing Cyber Ranges\\~\\ \small{[Short Discussion Paper]}}

\author{Lynsay A. Shepherd\inst{1}\orcidID{0000-0002-1082-1174} \and
Stefano De Paoli\inst{2}\orcidID{0000-0003-1120-4773} \and
Jim Conacher\inst{1}\orcidID{0000-0002-5712-0148}}

\authorrunning{L.A. Shepherd et al.}
\titlerunning{HCI Considerations When Developing Cyber Ranges}

\institute{Division of Cyber Security, School of Design and Informatics, Abertay University, Dundee, United Kingdom \and
Division of Sociology, School of Business, Law and Social Sciences, Abertay University, Dundee, United Kingdom\\
\email{\{lynsay.shepherd, s.depaoli, j.conacher\}@abertay.ac.uk}\\
}

\maketitle

%------------------------------------------------
%------------------------------------------------
%Abstract
\begin{abstract}
The number of cyber-attacks are continuing to rise globally. It is therefore vital for organisations to develop the necessary skills to secure their assets and to protect critical national infrastructure.  In this short paper, we outline upon human-computer interaction elements which should be considered when developing a cybersecurity training platform, in an effort to maintain levels of user engagement.  We provide an overview of existing training platforms before covering specialist cyber ranges.  Aspects of human-computer interaction are noted with regards to their relevance in the context of cyber ranges.  We conclude with design suggestions when developing a cyber range platform.

\keywords{Cybersecurity \and Security Awareness \and Cyber Range \and Human-Computer Interaction \and Cybersecurity Education.}
\end{abstract}
%------------------------------------------------
%------------------------------------------------
\section{Introduction}
\label{introduction}
In the field of cybersecurity, there is a growing interest in the design, development, and deployment of training platforms such as cyber ranges which can supplement and improve security professionals’ skills.  In this short paper, we aim to present an overview of existing cybersecurity training platforms, alongside a brief discussion of Human-Computer Interaction (HCI) elements which should be considered when developing a specialised cyber range platform.  We then offer guidance for improving and maintaining user engagement with these platforms through consideration of appropriate HCI techniques.

%------------------------------------------------
%------------------------------------------------
\section{Background}
\label{background}
This section provides a definition of HCI, and gives an overview of cybersecurity training platforms.  It then covers cyber ranges and their relevance in a secure modern society.

\subsection{Human-Computer Interaction}
Human-computer interaction (HCI) is a broad field which initially focused on a combination of human factors engineering and cognitive science \cite{carroll_hci}, and continues to link in with the areas of interaction design, ergonomics, informatics, and psychology.  HCI has also been incorporated into the field of cybersecurity, where it is termed HCISec (HCI security) and usable security.

Though HCI is linked to a number of fields and communities, the overarching goal is the \textit{``linkage of critical analysis of usability, broadly understood, with development of novel technology and applications"} \cite{carroll_hci}.

\subsection{Cybersecurity Training Platforms}
Training platforms are directly connected with the learning experience of the user; therefore, the user interface plays an essential role in both supporting learning pathways and keeping the users aware of the underlying processes simulated by the training platform.  

Cybersecurity training platforms have been used in a number of domains.  These educational tools range from small mobile applications geared towards raising public security awareness to those aimed at corporations.

Examples of such training platforms include:
\begin{itemize}
    \item Immersive Labs ``Human Cyber Readiness Platform" - aimed at businesses, and features hands-on scenarios with Capture The Flag Challenges \cite{ImmersiveLabs}.
    \item Cybersecurity Lab - browser-based game targeting young people to help them develop basic cyber security skills.  The user plays the role of a Chief Technology Officer who must defend a company against attacks \cite{PBS}.
    \item NoPhish - Android application to help the public identify phishing links \cite{canova2015nophish}.
\end{itemize}

\subsection{Cyber Ranges}
Cyber ranges can be defined as \textit{``interactive, simulated representations of an organization’s local network, system, tools, and applications that are connected to a simulated Internet level environment"} \cite{NIST2018}, and are a specific type of training platform created for security professionals.  They are typically composed of a virtual network environment and allow for the creation of simulated cyber-attack scenarios and incident response exercises.  There is a growing need for training platforms such as cyber ranges.  Owing to the sustained increase in cyber-attacks experienced by organizations around the World (particularly in the wake of the COVID-19 pandemic \cite{lallie2020cyber}), continually enhancing the cybersecurity resilience of such organizations is essential to help to ensure that critical national infrastructure remains protected.

Existing cyber ranges encompass a variety of areas, but they have generally been created for military, research and commercial purposes.  Examples of existing cyber ranges include the US Department of Defence Cyber Security Range \cite{MarinesRange} (military), the Austrian Institute of Technology Cyber Range \cite{AITRange} (academic) or the IBM Cyber Range \cite{IBMRange} (commercial).   Cyber ranges are a developing area for research e.g. the European Commission's H2020 Digital Security programme has funded platforms such as FORESIGHT \cite{FORESIGHT2019}.

%------------------------------------------------
%------------------------------------------------
\section{Discussion}
\label{discussion}
To ensure cyber ranges deliver an appropriate user experience in the context of an educational platform, we present design recommendations which aim to improve knowledge acquisition and maintain a high level of user engagement. 

\subsection{HCI and cyber ranges}
Although human-computer interaction is a large field, there are some key areas which are appropriate in the context of the cyber ranges.  This is not an exhaustive list of all applicable elements, but an overview of perhaps the most important aspects.  The areas mentioned offer the possibility of keeping the user engaged in the context of a cybersecurity training platform.

\subsubsection{User Interface (UI)}  The role of interface design in helping users learn has been explored in the context of e-learning.  Work by Guralnick \cite{guralnick2006user} highlights key factors in user interface design which aid the user.  These include the layout of elements on-screen (to guide the users' eye to look at the relevant information), consideration of learner paths to help the user stay focused, and well-presented guidance on-screen to provide the user with feedback.

Crucially, if the UI is difficult to navigate, the user will become frustrated, and this will detract from the learning process.  Existing cyber ranges such as the Kypo cyber range \cite{vceleda2015kypo} considered the role of the UI, and have utilised a portal based on Liferay Portals \cite{Liferay} to ensure users of all abilities can interact with the system. Developers should consider building upon existing frameworks to provide a suitable UI for a cyber range.

\subsubsection{Visualization} Information visualization has proved successful in supporting learning \cite{klerkx2014enhancing}.  Developers should consider deploying the use of user-centred design methods when creating visualisations in the cybersecurity domain \cite{mckenna2015unlocking}.  Many examples of cybersecurity visualisations already exist, including Kaspersky Cyber Threat map \cite{Kaspersky} and the Talos Spam and Malware Map \cite{Talos}.  Such tools could be incorporated into a cyber range to help the user assess the impact of potential threats e.g. identifying the source of a DDOS attack.

\subsubsection{Design Patterns}  Design patterns are design solutions to resolve common problems in software development.  These can utilise theories of motivation [p3] \cite{lewis2014irresistible} to create an engaging educational platform.  Additionally, these patterns can be designed to be gameful, linking in with section \ref{gamification} of this paper.  Gameful design patterns can incorporate some of the elements which are used in gamification, such as badges and leaderboards.  Gameful design patterns are particularly well-suited to applications with \textit{``heavy simulation elements that the user should explore"}, [p34] \cite{lewis2014irresistible} a category which cyber ranges fall into.

\subsubsection{Gamification}
\label{gamification}Gamification involves the use of gaming mechanics in traditionally non-gaming environments \cite{tondello2016introduction}.  Duolingo is a popular application which uses a combination of gamification elements such as learning paths, points, badges, scores, and leaderboards to help users learn new languages \cite{Duolingo}.

Gamification has been used in several existing cybersecurity training platforms and thus can be applied to cyber ranges.  Existing cybersecurity work which has utilised gamification includes prototype mobile applications aimed at raising public security awareness \cite{scholefield2019gamification}.  Furthermore, it has also been suggested for use in cyber defence training \cite{amorim2013gamified}, and used to tackle threats against critical national infrastructure \cite{cook2016using}. 
%------------------------------------------------
%------------------------------------------------
\section{Conclusion}
\label{conclusion}
In this paper, we have provided an overview of existing cybersecurity training platforms, and have highlighted the developing field of cyber ranges.  We have also outlined aspects of HCI which may help the end-user remain engaged with the platform, supporting learning and consolidating knowledge gained.  We hope that developers of cyber ranges will take these elements of human-computer interaction into consideration, creating an engaging cybersecurity platform.

%------------------------------------------------
%------------------------------------------------

%-----------------------
% ---- Bibliography ----
\bibliographystyle{splncs04}
\bibliography{references}

\end{document}